\documentclass[12pt,preprint]{aastex}

\shorttitle{$1_0-0_0$ transition of NH$_3$D$^+$}
\shortauthors{Dom\'enech et al.}

\begin{document}


\title{Improved determination of the 1$_{0}-0_{0}$ 
rotational frequency of NH$_{3}$D$^{+}$  from the high resolution 
spectrum of the $\nu_{4}$ infrared band}

\author{J. L. Dom\'enech, M. Cueto, V. J. Herrero \& I. Tanarro}
\affil{Molecular Physics Department, Instituto de Estructura de 
la Materia (IEM-CSIC). Serrano 123. 28006 Madrid, Spain}
\email{jl.domenech@csic.es}
\author{B. Tercero}
\affil{Department of Astrophysics, CAB. INTA-CSIC. Crta 
Torrej\'on-Ajalvir Km 4, 28850 Torrej\'on de Ardoz. Madrid, Spain}
\author{A. Fuente}
\affil{Observatorio Astron\'omico Nacional, Apdo. 112, 28803, Alcal\'a de Henares, Spain}
\and
\author{J. Cernicharo}
\affil{Department of Astrophysics, CAB. INTA-CSIC. Crta 
Torrej\'on-Ajalvir Km 4, 28850 Torrej\'on de Ardoz. Madrid, Spain}

\begin{abstract}

The high resolution spectrum of the $\nu_{4}$ band of NH$_{3}$D$^{+}$
 has been measured by difference frequency IR laser 
spectroscopy in a multipass hollow cathode discharge cell. From the set 
of molecular constants obtained from the analysis of the spectrum, a 
value of $262817\pm6$ MHz ($\pm3\sigma$) has been derived for the frequency of the 
1$_{0}-0_{0}$ rotational transition. This value supports the 
assignment to NH$_{3}$D$^{+}$ of lines at 262816.7 MHz recorded 
in radio astronomy observations in Orion-IRc2 and the cold prestellar 
core B1-bS.
\end{abstract}


\section{Introduction}

The concentrations of ions in interstellar space are typically orders of 
magnitude lower than those of the most abundant neutral species \citep{pet07, sno08, lar12} but, due to 
the fact that ion-molecule reactions are often barrier-less and have 
large rate coefficients, ions play a key role in the low-temperature gas 
phase chemistry of the interstellar medium (ISM) \citep{wat73, her73, her01}.
Many of the 20 cations detected up to date in the 
ISM \citep{pet07, lar12} are protonated 
derivatives of small molecules that can be formed in chains of proton 
transfer processes involving molecules with varying proton affinities 
(PA) \citep{bur70, car12a, car13}. 
The global initiator of these chains is assumed to be H$_3^+$,
 the most frequently produced ion in the ISM (Oka 2006), which is 
readily generated in the collisions of H$_{2}$ molecules with H$_{2}^{+}$ ions. 
Due to the low PA (422.3 kJ mol$^{-1}$) of the 
hydrogen molecule \citep{bur70, hun98}, H$_{3}^{+}$ 
has a high tendency to transfer a proton and return to H$_{2}$.
Among the small molecules most common in the ISM (CO, N$_{2}$, 
HCN, H$_{2}$O, NH$_{3}$, CH$_{2}$O, CH$_{4}$, CH$_{3
}$OH), ammonia has the highest proton affinity (PA=853 kJ mol$^{-1}$
) and consequently NH$_{4}^{+}$, once formed, will be stable 
toward further reactions with the most abundant species. In fact, 
astrochemical models predict NH$_{4}^{+}$ to become the dominant 
ion in some warm astronomical environments where NH$_{3}$ desorbs 
from interstellar grains and becomes available for gas phase 
ion-molecule chemistry \citep{rod03, har10, aik11}. It should be noted that, in spite of the 
concentration enhancement provided by grain desorption, NH$_{3}$ 
always represents a small fraction of the neutral species inventory. The 
dominance of NH$_{4}^{+}$ in the ion distributions of laboratory 
plasmas containing just small fractions of NH$_{3}$ has also been 
experimentally demonstrated \citep{car11, car13}. 

Interstellar NH$_{4}^{+}$ is not limited to warm environments. 
It is also assumed to play a key role in the low temperature gas-phase 
formation of ammonia in cold clouds \citep{nej90}. In this case, 
the suggested route starts with the reaction between N$^{+}$ and H$_{2}$ 
to yield NH$^{+}$ and proceeds through successive 
hydrogenation steps to the formation of NH$_{4}^{+}$ that 
transforms into NH$_{3}$ through dissociative recombination with 
electrons. The first step in this route is endoergic for para-hydrogen 
molecules in their ground state, but slightly exoergic for 
ortho-hydrogen, and \cite{dis12} have shown that realistic 
ortho/para ratios could explain the relative abundances of nitrogen 
hydrides observed in dark clouds. In general, the high gas-phase chemical 
stability of NH$_{4}^{+}$ in the ISM, which is mostly destroyed 
just in collisions with electrons, suggests that it can build up to high 
concentrations in different astronomical environments and, given its 
paramount importance for nitrogen chemistry in space, it is a desired 
goal for astronomical searches. Unfortunately, NH$_{4}^{+}$ is a 
spherical top molecule with no permanent electric dipole moment and thus 
unsuitable for radioastronomic observation. As far as we know, the ion 
has never been detected in the ISM. However, deuterated variants of NH$_4^+$ 
do possess a permanent dipole moment and could, in 
principle, be observed. Variously deuterated ammonia molecules, the 
likely precursors of deuterated ammonium, are observed in dense cores, 
where fractionation leads to an enhancement of the molecular deuterium 
content \citep{rou05} and, in particular, all deuterated 
isotopologues of NH$_{3}$ have been detected in the cold prestellar 
cloud B1-bS \citep{lis02}. 

Spectroscopic data useful for astronomic observations are available for 
the monodeuterated ammonium ion. Specifically, a frequency of $262807\pm9 $
MHz ($\pm3\sigma$) was predicted by \cite{nak86} for the 
1$_{0}-0_{0}$ transition of NH$_{3}$D$^{+}$ (as this paper will be 
often cited in this Letter, it will be abbreviated N\&A (1986) 
hereafter). This value was derived using the ground-state molecular 
constants obtained by those authors from their high resolution 
spectroscopic measurements of the $\nu_{4}$ infrared band of the ion. 
Recent work by Cernicharo et al. (2013, accompanying letter) has 
established the presence of an emission line at 262816.7 MHz in 
Orion-IRc2 and in B1-bS. The frequency is approximately 10 MHz higher 
than that estimated by Nagaka and Amano for monodeuterated ammonium and 
thus exceeds slightly the 3$\sigma$ uncertainty of the spectroscopic value, 
strongly suggesting that the feature can correspond to this species. In 
the present work we have refined the previous spectroscopic measurements 
of N\&A (1986) further constraining the frequency of the $1_0-0_0$
transition and providing additional support for the assignment 
of the observed 262816.7 MHz emission feature to the presence of NH$_{3}$D$^{+}$. 

\section{Experimental}

The experimental setup used in this work is similar to that described 
elsewhere \citep{tan94a, tan94b}. It consists of an IR 
difference-frequency laser spectrometer \citep{dom90, ber89},
and a hollow cathode discharge reactor refrigerated by 
room-temperature water, with multipass optics in a White cell 
configuration, similar to the design of \cite{fos84}. 

The IR radiation is generated by difference-frequency mixing, following 
the scheme of \cite{pin74}, of an Ar$^{+}$ laser and a tunable ring 
dye laser in a LiNbO$_3$ crystal. Tuning the dye laser, while keeping the 
Ar$^{+}$ laser frequency locked, tunes the IR frequency, which is the 
difference between that of the Ar$^{+}$ laser and that of the ring 
dye laser. In our experiment, the $\rm Ar^+$ laser is frequency locked 
to a hyperfine transition of $^{127}$I$_{2}$, (a$_{3}$ 
component of the P(13) $43-0$ transition, known with an accuracy $\sim0.1$ 
MHz, \citealt{qui03}) and has a residual frequency jitter less than 1 MHz 
and long-term stability of the same order of magnitude. The wavelength 
of the ring dye laser (also frequency stabilized, with a residual jitter 
of $\sim3$ MHz) is measured in just 1 ms at each data point of the spectrum 
by a  high accuracy wavemeter based on Fizeau interferometers (High Finesse WSU10) with a 
manufacturer-stated absolute accuracy of better than 10 MHz (3$\sigma$). 
The $\rm Ar^+$ laser is used to calibrate the wavemeter and thus insure its 
accuracy throughout the experiments. The linewidth of the IR beam is the 
combined linewidth of those of the visible lasers, i.e. $\sim3$ MHz. The 
internal coherence of the infrared frequency scale is limited by that of 
the wavemeter. From repeated measurements of N$_{2}$O lines in the 
$3300-3400$ cm$^{-1}$ region, and their comparison with the values in 
the HITRAN 2008 database \citep{rot09, tot04}, we have 
verified a reproducibility and internal coherence of our IR frequency 
scale better than 1.2$\times $10$^{-4}$ cm$^{-1}$ (4 MHz) rms and 
a systematic offset of 6$\times $10$^{-4}$ cm$^{-1}$ (18 MHz), 
which is within the combined uncertainty of the wavemeter and the quoted 
uncertainty limits for those line frequencies in HITRAN ((1-10)$\times10^{-4}$ cm$^{-1}$, or $3-30$ MHz).

The IR beam traverses the multipass cell 12 times, giving an effective 
absorption length of 9 meters inside the discharge. In order to improve 
the detection sensitivity, both the IR beam and the discharge are 
amplitude-modulated, at 14.2 and 5.5 kHz, respectively, and the IR 
signal is detected by an InSb detector connected to a dual-phase lock-in 
whose reference frequency is the sum of the other two (i.e. 19.7 kHz). 
In this way, only changes in the IR beam intensity, due to species whose 
concentration is modulated in the discharge, are detected. The three 
frequencies are derived from a common oscillator by appropriately chosen 
integer dividers \citep{dom94}, therefore, insuring phase 
stability between the three of them. The detection time constant is 100 
ms with 12 db/oct roll off. The ring dye laser scans at 0.005 cm$^{-1
}$s$^{-1}$ rate. Each line has been recorded 10 times and the results 
averaged in a grid spaced 0.001 cm$^{-1}$.

The discharge is modulated at 5.5 kHz through an audio amplifier, a 
step-up transformer and a circuit designed to avoid the transformer core 
saturation, since the hollow cathode discharge acts inherently as an 
electric rectifier. Typical discharge conditions are 200 mA and 400 V 
rms. The relatively high modulation frequency has been chosen to favor 
the detection of ionic species, whose lifecycles follow closely the 
on-off condition of the discharge, and to minimize undesired absorption 
signals associated with much slower processes like concentration change 
of the neutral species caused by multistep wall reactions \citep{car12b, car13}.
Nevertheless, we have observed spectral signatures 
from NH$_{3}$ due to a combination of population and Doppler width 
changes that occur with the small temperature change on the "on" 
periods of the discharge, similar to those observed in \cite{tan94a}
for other stable species. Note that the kinetic temperature 
deduced from the observed Doppler widths of the ions is $\sim315$ K, i.e. 
very near the temperature of the refrigerated cathode.

A 32 Pa (0.32 mbar) gas flowing mixture of NH$_{3}$ (37\%) + D$_{2}$ (63\%) is 
used in the discharge as plasma precursor. Gas flows, measured with 
rotameters at $5\times10^{4}$ Pa (500 mbar) absolute input pressure, were 15 ml/min and 30 
ml/min for NH$_{3}$ and D$_{2}$, respectively. In the hollow cathode 
discharges used, efficient atomic recycling at the metallic cathode wall 
\citep{jim11, car11} leads to an 
extensive isotopic scrambling both in the neutrals and ions. Most of the 
initial NH$_{3}$ is transformed into N$_{2}$ and H$_{2}$ and 
the pressure raises to $\sim46$ Pa ($\sim0.46$ mbar). The mixture proportions, flow rate and 
pressure were empirically selected to maximize the NH$_{3}$D$^{+}$ concentration.

Although the ion concentrations have not been directly measured in the 
present experiment, previous kinetic studies and Langmuir probe 
measurements by our group \citep{car11, car13} 
on comparable hollow cathode discharges of H$_{2}$/N$_{2}$ 
mixtures, indicate that the ion distributions in our cell should be 
dominated by NH$_{4}^{+}$, with smaller concentrations of N$_{2}$H$^{+}$ 
and H$_{3}^{+}$ (and their deuterated variants) 
and that the total ion densities should be of the order of 
$\sim(1-5)\times $10$^{10}$ cm$^{-3}$, with electron temperatures 
of $\sim3.0\pm0.5$ eV. These density values are also consistent with 
estimates made from the measured cathode current and glow region 
dimensions using Bohm's velocity \citep{lie94}. The 
concentrations of the main individual ions and, in particular, of NH$_{
3}$D$^{+}$ are estimated to be in the 10$^{9}-10^{10}$ cm
$^{-3}$ range.

It must be noted that this set-up is very similar to the one used by 
N\&A (1986). The main differences between them are: the IR frequency 
measuring system (a high accuracy wavemeter for the dye laser in our 
case, and calibration with N$_{2}$O IR absorption lines with 
1$\times $10$^{-3}$ cm$^{-1}$ (30 MHz) accuracy in the previous 
work), the detection scheme at the sum frequency vs. detection at the 
discharge frequency, the different frequencies of the discharge (5.5 kHz 
vs. 17 kHz), the rather different proportion of NH$_{3}$ and D$_{2}$
in the gas mixture ($\sim1:2$ in our case vs. 1:10 in the previous work), 
the different absorption path lengths (9 m in this work vs. 20 m in the 
previous one) and the IR power available ($\sim1$ $\mu$W in this work vs. $\sim10$ 
$\mu$W in the previous one). We would like to emphasize that our experiment 
has been guided and facilitated extensively by the previous study of 
N\&A (1986). The major experimental improvement of the present study is due to the 
availability of a more precise frequency scale and a more favorable 
ratio of precursors in the discharge.

\section{Results and discussion}

We have recorded 114 lines between 3268.4 and 3414.7 cm$^{-1}$. Not 
all the spectral interval has been scanned. Instead, we have chosen to 
scan selected regions around the line frequencies calculated with the 
constants provided by N\&A (1986). In each region, 10 recordings have 
been made and averaged, in order to improve the signal to noise ratio. 
From the repeatability of the shape of the most intense lines, and from 
the fact that no broadening is observed in the averages, we are assured 
that the frequency scale derived from the wavemeter is highly 
reproducible. The predictions, new assignments and fits have been made 
with the help of the PGOPHER program \citep{wes10}. A $\sim1$ cm$^{-1}$
wide region of the spectrum in the central part of the band is shown in 
Figure 1, which can be compared with Figure 1 of N\&A (1986). The more 
prominent features are due to NH$_4^+$, which we have 
identified throughout the spectrum with the measurements and predictions 
of \cite{sch84}. Our much favorable intensity ratio between 
NH$_3$D$^+$ and NH$_{4}^{+}$ signals as compared with 
those of N\&A (1986) is worth noting. Figure 2 shows some more examples 
of more isolated lines of NH$_{3}$D$^{+}$ in the {\it P-} and {\it R-} 
regions. 

Each averaged line has been fit to a Gaussian function, to derive its 
line center. Standard deviations of the center frequency derived from 
these fits are $(1-4)\times10^{-4 }$ cm$^{-1}$ ($3-12$ MHz). 
The line center frequencies have been fit to the same Hamiltonian used 
in the work of N\&A (1986), with the only exception of the sign of the 
$q_+$ (rotational $l$-doubling) parameter, since PGOPHER uses the 
opposite sign convention for it in its matrix elements.  For completeness,
we give here the expressions used for the energy levels, 
which are common for a prolate symmetric-top.  For the nondegenerate ground-state, ignoring spin-rotation effects:
\begin{equation}
\begin{array}{l}
 E''(J,K)= \\
 B''J(J+1) + (A''-B'')K^2 \\
- \mbox{ }D''_JJ^2(J+1)^2 - D''_{JK}J(J+1)K^2 - D''_KK^4  \\
 + \mbox{ higher order centrifugal distortion terms,}
\end{array}
\end{equation}
and for the degenerate $v_4=1$ state
\begin{equation}
\begin{array}{l}
E'(J,k) = \nu_0\\
+\mbox{ }  B'J(J+1) + (A'-B')k^2 \\
 + \mbox{ }(-2A'\zeta + \eta_JJ^2 + \eta_Kk^2) lk \\
 -\mbox{ } D'_JJ^2(J+1)^2 - D'_{JK}J(J+1)k^2 - D_Kk^4 \\
 + \mbox{higher order centrifugal distortion terms,}
\end{array}
\end{equation}
with the off-diagonal matrix elements, responsible for the rotational
$l$-doubling,
\begin{equation}
\begin{array}{l}
\langle J,k+2,l+1|H|J,k,l-1\rangle = \frac{1}{2}q_+
[(J(J+1)\\
\mbox{ }-k(k-1))(J(J+1)-k(k+1))]^{\frac{1}{2}}
\end{array}
\end{equation}
$A$ and $B$ are the rotational constants perpendicular and parallel, respectively, to the symmetry axis.  The $D$ constants account for the centrifugal distortion effects.  $J$ and $K$ are the usual rotational quantum numbers, with $K=|k|$.  Recall that Coriolis interaction splits $k\ne0$ levels in the degenerate vibrational state into $+l$ 
and $-l$ levels  (involving the parameters $\zeta$, $\eta_J$ and 
$\eta_K$ above) and that the $(\Delta k=\pm2, \Delta l=\pm2)$ interaction
splits the $k=l=\pm1$ levels into pairs (involving the $q_+$ parameter).
For a more detailed description of the vibration-rotation Hamiltonian 
and the matrix elements used here the reader is referred to \cite{pap82} and PGOPHER's documentation \citep{wes10}, respectively.

Finally, 76 lines have been fitted, 15 more than in the N\&A (1986) 
work. The lines that have not been included in the fit either show 
interference from NH$_{4}^{+}$ or NH$_{3}$ features, have 
very poor signal to noise ratio, or show abnormally high linewidth 
(indicating a possible overlap). Besides, we have restricted ourselves 
mostly to lines with $J, K \le 7$, in order to avoid using sextic and higher 
centrifugal distortion constants. Restriction to these quantum numbers, 
even at the expense of reducing the number of observations included in 
the fit, provides a better description of the low lying levels that we 
pursue, avoiding the correlations between parameters that appear as the 
number of degrees of freedom is increased without a much larger number 
of observations.

Table 1 shows the observed frequencies, observed minus calculated 
residuals and the assignments of the fitted lines. The standard 
deviation of the fit is 5$\times $10$^{-4}$ cm$^{-1}$. Table 2 
shows the resulting constants and their standard deviations (1$\sigma$) derived 
from the fit. Since $\nu_{4}$ is a perpendicular band of a symmetric 
top, $A'', A', D''_{K}, D'_K$ and $\zeta$ cannot be independently 
determined from the observations, so $A''$ and $D''_{K}$ have been 
constrained to the same values as in the N\&A (1986) paper. As indicated 
there, this choice of parameters affects $\nu_{0}$, $\zeta$, A$'$ and 
$D'_K$ values, while the rest remain unaffected. Specifically, the 
frequency of the $J_K=1_0-0_0$ pure rotational 
transition, namely $2B''-4D''$, does not depend on any of these parameters, 
so our estimation is not influenced by the constraints. Table 2 also 
shows the results from N\&A (1986), showing that the most significant 
changes are in the $B$ and $D$ rotational constants. From the $B''$ and 
$D''_J$ values, we obtain a frequency for the $1_0-0_0$ 
transition of 262816.8 MHz, in good agreement with the 
observation of Cernicharo et al. (2013, accompanying letter). We are aware 
that the near perfect match is probably fortuitous, given the precision 
of the individual IR frequencies and the statistical uncertainty of the 
parameters derived from the fit.

In order to summarize, based on our observations and the analysis 
described above, we propose a value of $262817\pm6$ MHz ($\pm3\sigma$) for the 
frequency of the 1$_{0}-0_{0}$ pure rotational transition of NH$_{3}$D$^{+}$. 
Our IR measurements improve upon those of N\&A 
(1986) thanks to a more precise frequency scale and a larger number of 
observed lines. This frequency is in good agreement with that of the 
feature detected in Orion-IRc2 and B1-bS (Cernicharo el al. 2013, 
accompanying letter), thus supporting the identification of NH$_{3}$D$^{+}$ in the ISM.

\acknowledgements

The authors acknowledge the support from the Spanish MICINN through 
grant CSD2009-00038. JLD acknowledges additional financial support 
through grant FIS2012-38175; VJH and IT acknowledge additional financial 
support through grant FIS2010-16455; BT and JC acknowledge additional 
financial support through grant AYA2009-07304. JC thanks U. Paris Est for 
an invited professor position during the completion of this work. Our skilful technicians 
M. A. Moreno, and J. R. Rodr\'{i}guez are gratefully acknowledged.

\begin{deluxetable}{crccrcc}
\tabletypesize{\small}
\tablewidth{0pt}

\tablecaption{Fitted lines, residuals after the fit and the quantum numbers assignment.}

\tablenum{1}

\tablehead{\colhead{$\nu_{\rm{obs}}/\rm{cm}^{-1}$} & \colhead{(o-c)\tablenotemark{a}} & 
\colhead{$J'$} & \colhead{$K'$} & \colhead{$l$} & \colhead{$J''$} & \colhead{$K''$} } 

\startdata
3275.00762 & 1.2   & 5 & 5 &  -1  & 6 & 6 \\
3277.36239 & 6.3   & 5 & 4 &  -1  & 6 & 5 \\
3279.75397 & 4.3   & 6 & 1 &  1   & 7 & 0 \\
3281.97669 & 1.2   & 5 & 2 &  -1  & 6 & 3 \\
3284.24462 & -0.6  & 5 & 1 &  -1  & 6 & 2 \\
3286.24996 & -7.2  & 4 & 4 &  -1  & 5 & 5 \\
3286.48992 & -2.0  & 5 & 0 &      & 6 & 1 \\
3288.58604 & 4.2   & 4 & 3 &  -1  & 5 & 4 \\
3288.71980 & 3.2   & 5 & 1 &  1   & 6 & 0 \\
3290.89042 & -1.4  & 4 & 2 &  -1  & 5 & 3 \\
3293.16846 & -3.7  & 4 & 1 &  -1  & 5 & 2 \\
3297.46387 & 0.7   & 3 & 3 &  -1  & 4 & 4 \\
3299.77847 & -0.8  & 3 & 2 &  -1  & 4 & 3 \\
3302.06568 & 4.4   & 3 & 1 &  -1  & 4 & 2 \\
3306.56586 & 6.1   & 3 & 1 &  1   & 4 & 0 \\
3310.93202 & 2.8   & 2 & 1 &  -1  & 3 & 2 \\
3313.19858 & 2.9   & 2 & 0 &      & 3 & 1 \\
3315.44070 & 1.7   & 2 & 1 &  1   & 3 & 0 \\
3319.76669 & 3.7   & 1 & 1 &  -1  & 2 & 2 \\
3322.03737 & 5.4   & 1 & 0 &      & 2 & 1 \\
3330.84074 & 6.5   & 0 & 0 &      & 1 & 1 \\
3334.30357 & -7.6  & 6 & 2 &  -1  & 6 & 3 \\
3334.68004 & -9.7  & 4 & 2 &  -1  & 4 & 3 \\
3334.81962 & -1.3  & 3 & 2 &  -1  & 3 & 3 \\
3336.32317 & 5.3   & 7 & 1 &  -1  & 7 & 2 \\
3336.57102 & 2.0   & 6 & 1 &  -1  & 6 & 2 \\
3336.96581 & -8.1  & 4 & 1 &  -1  & 4 & 2 \\
3337.11283 & 5.2   & 3 & 1 &  -1  & 3 & 2 \\
3338.81264 & 3.3   & 6 & 0 &      & 6 & 1 \\
3339.03585 & -7.1  & 5 & 0 &      & 5 & 1 \\
3339.22520 & 1.5   & 4 & 0 &      & 4 & 1 \\
3339.37658 & -3.1  & 3 & 0 &      & 3 & 1 \\
3339.49136 & -0.3  & 2 & 0 &      & 2 & 1 \\
3339.56780 & -2.2  & 1 & 0 &      & 1 & 1 \\
3341.02311 & -0.8  & 6 & 1 &  1   & 6 & 0 \\
3341.25759 & 0.8   & 5 & 1 &  1   & 5 & 0 \\
3341.45410 & -2.9  & 4 & 1 &  1   & 4 & 0 \\
3341.61294 & -0.1  & 3 & 1 &  1   & 3 & 0 \\
3341.73208 & -4.0  & 2 & 1 &  1   & 2 & 0 \\
3342.63625 & -6.3  & 8 & 2 &  1   & 8 & 1 \\
3342.94920 & 0.8   & 7 & 2 &  1   & 7 & 1 \\
3343.22602 & 6.0   & 6 & 2 &  1   & 6 & 1 \\
3343.46451 & -0.2  & 5 & 2 &  1   & 5 & 1 \\
3343.66511 & -3.0  & 4 & 2 &  1   & 4 & 1 \\
3343.82666 & -5.2  & 3 & 2 &  1   & 3 & 1 \\
3345.11471 & 3.2   & 7 & 3 &  1   & 7 & 2 \\
3345.39715 & -1.2  & 6 & 3 &  1   & 6 & 2 \\
3345.64164 & -4.0  & 5 & 3 &  1   & 5 & 2 \\
3345.84661 & -10.1 & 4 & 3 &  1   & 4 & 2 \\
3346.01221 & -9.4  & 3 & 3 &  1   & 3 & 2 \\
3347.25607 & -5.2  & 7 & 4 &  1   & 7 & 3 \\
3347.54483 & 1.6   & 6 & 4 &  1   & 6 & 3 \\
3347.79345 & -4.4  & 5 & 4 &  1   & 5 & 3 \\
3348.00364 & 4.6   & 4 & 4 &  1   & 4 & 3 \\
3350.57916 & -6.1  & 1 & 1 &  1   & 0 & 0 \\
3359.26698 & 6.9   & 2 & 1 &  1   & 1 & 0 \\
3361.47942 & -8.8  & 2 & 2 &  1   & 1 & 1 \\
3367.91091 & -0.9  & 3 & 1 &  1   & 2 & 0 \\
3370.12020  & -0.8 & 3 & 2 &  1   & 2 & 1 \\
3372.30329 & -3.2  & 3 & 3 &  1   & 2 & 2 \\
3376.51343 & 4.9   & 4 & 1 &  1   & 3 & 0 \\
3378.71679 & 7.8   & 4 & 2 &  1   & 3 & 1 \\
3380.89518 & 4.8   & 4 & 3 &  1   & 3 & 2 \\
3383.04468 & 3.0   & 4 & 4 &  1   & 3 & 3 \\
3385.07123 & -1.3  & 5 & 1 &  1   & 4 & 0 \\
3387.26751 & 7.7   & 5 & 2 &  1   & 4 & 1 \\
3389.44028 & 4.5   & 5 & 3 &  1   & 4 & 2 \\
3391.58446 & 1.2   & 5 & 4 &  1   & 4 & 3 \\
3393.58543 & -2.2  & 6 & 1 &  1   & 5 & 0 \\
3393.69878 & 6.8   & 5 & 5 &  1   & 4 & 4 \\
3400.07709 & 0.8   & 6 & 4 &  1   & 5 & 3 \\
3402.05462 & -7.8  & 7 & 1 &  1   & 6 & 0 \\
3408.26843 & 5.6   & 8 & 0 &      & 7 & 1 \\
3408.52168 & -3.1  & 7 & 4 &  1   & 6 & 3 \\
3410.62517 & 5.8   & 7 & 5 &  1   & 6 & 4 \\
3414.72838 & -2.9  & 7 & 7 &  1   & 6 & 6 \\
\enddata                         
\tablenotetext{a}{(o-c)=$(\nu_{\rm{obs}}-\nu_{\rm{calc}})/10^{-4}$ cm$^{-1}$}

\end{deluxetable}

\begin{deluxetable}{lrr}
\tablewidth{0pt}

\tablecaption{Constants derived from the fit}
\tablenum{2}

\tablehead{\colhead{Constants} & \colhead{This work} & \colhead{N\&A (1986)} \\ 
\colhead{$/$cm$^{-1}$} & \colhead{} & \colhead{} } 

\startdata
$A''$	      & 5.852\tablenotemark{a}	         & 5.852\tablenotemark{a}	 \\
$B''$	      & 4.3834351(294)		                & 4.38327(5)	 \\
$D''_J$	    & 6.1363(373)$\times10^{-5}$	      &	5.87(9)$\times10^{-5}$	 \\
$D''_{JK}$  & 1.4689(293)$\times10^{-4}$        &1.52(6)$\times10^{-4}$ \\
$D_K''$	    & 0.0\tablenotemark{a}		          & 0.0\tablenotemark{a}	 \\
$\nu_0$	    & 3341.07498(17)		                & 3341.0764(3)	 \\
$A'$	      &   5.818834(37)		                & 5.81871(9)	 \\
$B'$	      &   4.3640729(278)		              & 4.36391(5)	 \\
$D'_J$	    & 5.4024(339)$\times10^{-5}$		    & 5.13(10)$\times10^{-5}$	 \\
$D_{JK}'$   & 9.705(296)$\times10^{-5}$		      & 1.02(7)$\times10^{-4}$	 \\
$D_K'$	    & 3.801(91)$\times10^{-5}$		      & 3.1(3)$\times10^{-5}$	 \\
$\zeta$	    & 0.0582020(76)		                  & 0.058191(14)\tablenotemark{b}	 \\
$\eta_J$    & -4.2581(686)$\times10^{-4}$	      & -4.23(13)$\times10^{-4}$	 \\
$\eta_K$    &	1.744(74)$\times10^{-4}$		      & 1.76(18)$\times10^{-4}$	 \\
$q_+$	      & -3.393(98)$\times10^{-4}$\tablenotemark{c}		        & 2.93(19)$\times10^{-4}$	 \\
\enddata

\tablenotetext{a}{Constrained in the fit}
\tablenotetext{b}{Calculated from the values of A$'$ and A$'\zeta$ given in N\&A (1986)}
\tablenotetext{c}{Sign convention is opposite to that of N\&A (1986)}
\tablecomments{Numbers in parentheses are one standard deviation in units of the last quoted digit, 
as derived from the fit.  For this work we give all the significant digits necessary to reproduce 
the calculated line frequency values.}

\end{deluxetable}

\begin{figure}
\includegraphics[angle=0,scale=.4]{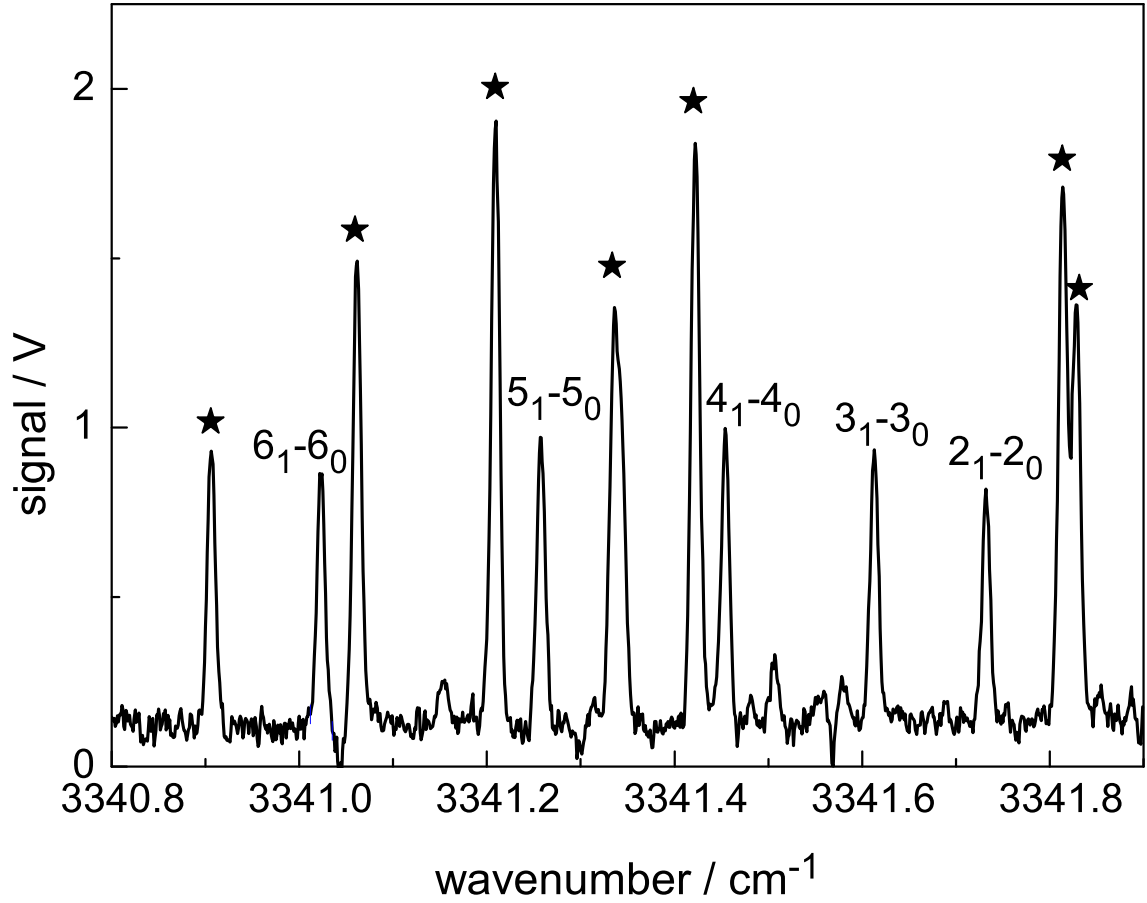}
\caption{Portion of the $\nu_4$ band of NH$_3$D$^+$ near the band center.  
The labeled lines belong to the $(J, K'=1)\leftarrow(J,K''=0)$ 
progression ($^rQ_0(J)$ in branch notation), and are the same as those 
shown in Figure 1 of N\&A (1986).  Lines marked with an asterisk belong to NH$_4^+$..}
\end{figure}

\begin{figure}
\includegraphics[angle=0,scale=.5]{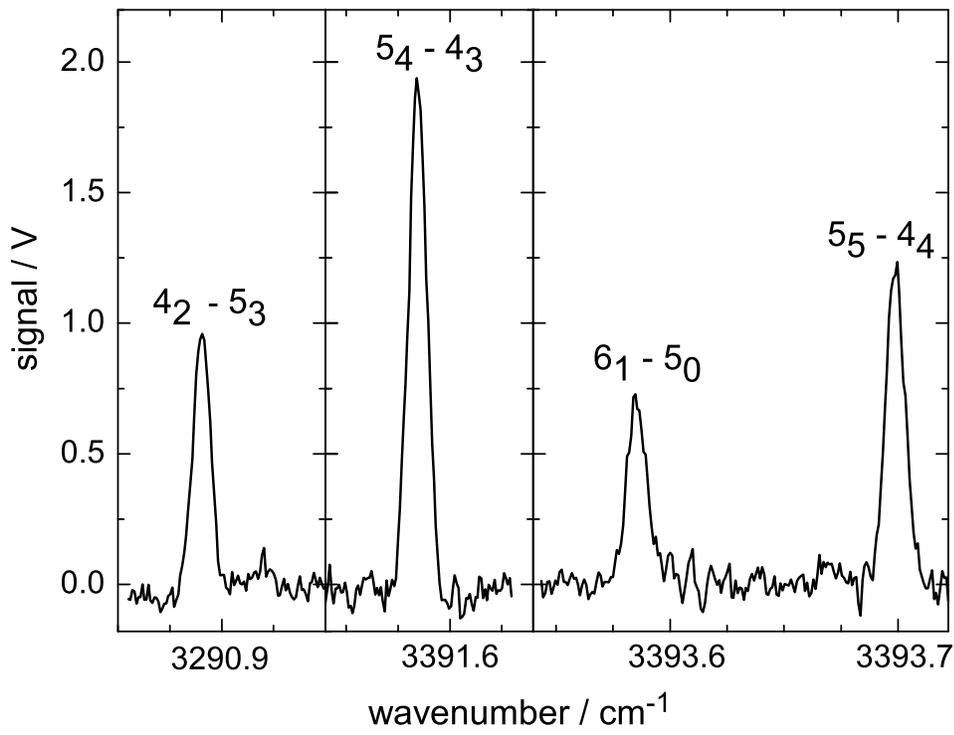}
\caption{Additional examples of $\nu_4$ lines recorded in this work.  
One horizontal division is 0.025 cm$^{-1}$.}
\end{figure}

\end{document}